\documentclass{elsarticle}

\usepackage{tikz}
\usetikzlibrary{arrows}
\usetikzlibrary{decorations.markings}
\tikzset{
o/.style={
shorten >=#1,
decoration={
markings,
mark={
at position 1
with {
\draw[fill=black] circle [radius=#1];
}
}
},
postaction=decorate
},
o/.default=2pt
}
\usetikzlibrary{calc}

\newcommand{\vect}[1]{\mathbf{#1}}

\usepackage{rotating}
\usepackage{amsmath,amssymb}
\usepackage{todonotes}
\usepackage{url}

\begin{document}

\begin{frontmatter}
\date{}
\title{A biological model of scabies infection dynamics and treatment explains why mass drug administration does not lead to elimination.}

\author[uom,mcri]{M.~Lydeamore}
\author[doherty,uom2,mcri]{P.T.~Campbell}
\author[kirby]{D.G.~Regan}
\author[doherty,menzies]{S.Y.C.~Tong}
\author[menzies]{R.~Andrews}
\author[mcri]{A.C.~Steer}
\author[kirby]{L.~Romani}
\author[kirby]{J.M.~Kaldor}
\author[doherty,uom2,mcri]{J.~McVernon}
\author[uom,uom2,mcri]{J.M.~McCaw\corref{cor1}}
\ead{jamesm@unimelb.edu.au}
\cortext[cor1]{Corresponding author}

\address[uom]{School of Mathematics and Statistics, The University of Melbourne}
\address[mcri]{Murdoch Childrens Research Institute, The Royal Children's Hospital, Melbourne}
\address[uom2]{Melbourne School of Population and Global Health, The University of Melbourne}
\address[kirby]{Kirby Institute, UNSW Australia}
\address[menzies]{Menzies School of Health Research, Charles Darwin University}
\address[doherty]{Peter Doherty Institute for Infection and Immunity, The Royal Melbourne Hospital and The University of Melbourne}

\begin{abstract}
Despite a low global prevalence, infections with \emph{Sarcoptes scabiei}, or scabies, are still common in remote communities such as in northern Australia and the Solomon Islands. Mass drug administration (MDA) has been utilised in these communities, and although prevalence drops substantially initially, these reductions have not been sustained. We develop a compartmental model of scabies infection dynamics and incorporate both ovicidal and non-ovicidal treatment regimes. By including the dynamics of mass drug administration, we are able to reproduce the phenomena of an initial reduction in prevalence, followed by the recrudescence of infection levels in the population. We show that even under a `perfect' two-round MDA, eradication of scabies under a non-ovicidal treatment scheme is almost impossible. We then go on to consider how the probability of elimination varies with the number of treatment rounds delivered in an MDA. We find that even with infeasibly large numbers of treatment rounds, elimination remains challenging.
\end{abstract}
\end{frontmatter}

\section{Introduction}
Infections with the mite \emph{Sarcoptes scabiei}, commonly known as scabies, are relative uncommon in urban, well-developed environments. However, in many lower income settings, particularly in tropical regions, scabies remains endemic. In remote communities in northern Australia, for example, prevalence is as high as $49\%$, and in the Solomon Islands and Fiji, prevalence is $43\%$ and $28\%$, respectively\cite{Romani2015a}. Scabies is highly contagious, and causes intense itching on the host \cite{Mccarthy2004}.  Besides the psychological impact due to the constant itching \cite{Hotez2014}, the scratching leads to a break in the skin layer, creating a pathway for secondary skin infections such as Group A \emph{Streptococcus} (GAS) to take hold \cite{Currie2000}. It has been hypothesized that controlling scabies infections could lead to a reduction in the disease burden attributable to GAS and its sequelae \cite{Currie2000}. However, despite multiple trials confirming the short term effectiveness of scabicidal therapies, follow up studies in several communities have shown recrudescence of infection within months to years of treatment cessation \cite{LaVincente2009, Andrews2009, Wong2002,Kearns2015}.

Mathematical models provide useful frameworks in which to consider the drivers of infectious disease, with a view to optimising treatment approaches. To our knowledge, there exist only two models for scabies infection in humans \cite{Bhunu2013,Gilmore2011}, and neither of these models attempts to capture the natural history of the mite's life cycle in relation to the host. This omission is important in understanding intervention effects, as the parasite's life state can interact critically with treatment success or failure.

Here, we develop a model of scabies infection and use it to explore the likely impact of mass drug administration treatment strategies. The structure of this paper is as follows: In Section 2, we summarise the biology of the mite and the effect of ovicidal and non-ovicidal treatments. In Section 3, we develop and introduce a compartmental model for scabies, including the effects of different treatment mechanisms. In Section 4, the results of the investigation into the model are presented, and in Section 5, the implications of our investigation are discussed and summarised.

\section{Scabies Biology and Treatment} \label{sec:biology}
The scabies mite progresses through three general life stages: egg, young mite and adult. The eggs are relatively well studied, and are believed to take approximately two days to hatch \cite{Arlian1988,Walton2004,Currie2010}. The young mite stage is more complex, comprising a number of developmental stages. Initially, mites are considered larvae, and are unlikely to emerge from the burrow in which the eggs were laid. Mites remain as larvae for approximately five days \cite{Arlian1988,Walton2004}, before developing into Protonymphs and Tritonymphs \cite{Arlian1988,Currie2010}. In both of the nymph stages, the mites roam about the body \cite{Mellanby1944}. Finally, the nymphs develop into adult mites, form breeding pairs, and the pregnant female mite lays eggs. The second generation of adult mites appear after approximately 30 days of initial infestation \cite{Mellanby1944}. As it takes approximately five days for the nymphs to become adults \cite{Arlian1988,Currie2010}, it follows that it must take approximately two weeks for an adult mite to find a mate.

Consider the infestation cycle of an individual human host. An individual is initially infected with an already impregnated mite. The pregnant mites begin tunneling almost immediately once transferred \cite{UniversityofSydney2016}, and lay 2-3 eggs per day. These eggs hatch, and eventually develop into adult mites, completing the cycle of infestation.

In an individual's first infection, a long asymptomatic phase is experienced, lasting for up to 60 days \cite{Mellanby1944}. In subsequent infections, the onset of itching is almost instantaneous due to prior sensitisation, and the mite count is demonstrably lower.

In the absence of treatment, little is known about natural recovery. In human experiments, however, no individuals saw natural recovery after being infected for almost 200 days \cite{Mellanby1944}.

As the infestation can only begin with a pregnant mite (we discount the logical possibility that the host has a male \emph{and} non-pregnant female transferred to them \emph{and} that they subsequently mate), we assume that an individual is only infectious if they are  harboring pregnant mites.

There are three widely-used treatments for scabies: Permethrin, Benzyl Benzoate and Ivermectin. Permethrin and Benzyl Benzoate are a topical creams which must be applied to the whole body for at least eight hours, a requirement associated with poor compliance \cite{Usha2000}. Ivermectin is administered as a single oral dose, leading to improved compliance. A key difference between the treatments is that Permethrin and Benzyl Benzoate are believed to be ovicidal (egg-killing), and thus only one treatment may be necessary for clearance \cite{Currie2000,Usha2000}. In contrast, Ivermectin is believed to be non-ovicidal, and so at least two treatments are required to eliminate all the stages of the mite. In fact, two doses of Ivermectin have been shown to significantly increase the probability of clearance, when compared with a single dose \cite{Usha2000}. Generally, in the absence of a second non-ovicidal treatment, \emph{endogenous} or continuing self-reinfection is inevitable.

Consider again the infestation cycle of an individual. In the event of an infested individual receiving a $100\%$ effective non-ovicidal treatment, only the eggs will remain on the individual. The eggs will inevitably hatch into larvae, and, in the absence of a second treatment, eventually develop into adults, continuing the cycle of infestation. To clear the infestation successfully, a second treatment will be required before the new generation of hatched mites begins to lay eggs.

Although no natural recovery from primary scabies infections was observed in human studies, a reduced level of infestation was observed in subsequent infections, suggestive of some degree of immune-mediated suppression \cite{Mellanby1944}. Moreover, in `real world' settings, continuous background treatment of scabies occurs through public health clinics as cases are identified. In areas with endemic infection, this background treatment may occasionally be supplemented through a \emph{mass drug administration} (MDA) using either an ovicidal or non-ovicidal treatment regime.

\section{Model Development} \label{sec:markovmodel}
We introduce a compartmental mathematical model to characterise scabies transmission and treatment in a population with high endemic prevalence, and capture the potential differences between ovicidal and non-ovicidal treatments. First, we develop the model considering the non-ovicidal treatment regime, which does not kill the eggs laid by the mite. Later we will consider the model including ovicidal treatment.

In addition to susceptible and infectious states, a population level model of scabies with sufficient biological fidelity to study how an MDA impacts upon the population must also consider a number of other states. As non-ovicidal treatment for scabies kills living mites, but not the eggs of the mite, it is essential to keep track of the the proportion of the population with only eggs, as in the absence of a second treatment course, endogenous reinfection upon hatching is inevitable. Given that the maturation time of the mite is notably longer than the amount of time it takes the eggs to hatch, these life stages can be considered mutually exclusive. As such, individuals in a population can be broadly categorised into one of four mutually exclusive states: Susceptible $(S)$, Infectious $(I)$, Infectious and with eggs present $(\hat{I})$, and having only eggs $(\hat{G})$. Throughout, we use a hat ( $\hat{}$ ) to signify states with eggs present. A non-ovicidal treatment will move an infectious individual from $I$ to $S$, or from $\hat{I}$ to $\hat{G}$, depending on whether or not the individual currently has eggs present. Left untreated, all modelled individuals will eventually reside in the $\hat{I}$ class, having both living mites and eggs. The transitions of this system as a Markov chain are given in Equations (\ref{eqn:model1eqn1}-\ref{eqn:model1eqnlast}), and are represented by Figure \ref{fig:model1}. This model accounts for eggs using the $\hat{I}$ and $\hat{G}$ states, and also accounts for the fact that non-ovicidal treatment only kills the live mites, leaving eggs.

However, this model does not explicitly account for the sexual maturation of the mite. This omission leads to two issues: firstly, the model implies that as soon as any mite hatches on an individual who was only harboring eggs, the individual is now harboring a mature, fertile adult mite. In reality, this is not the case, as newborn mites undergo a period of sexual maturation and development. Secondly, as an individual is harboring pregnant mites as soon as an egg hatches, the period of time until new eggs are laid is small, and so the likelihood of a successful second treatment resulting in total clearance of the mites is negligible.

\tikzstyle{int}=[draw, minimum size=2em]
\tikzstyle{init} = [pin edge={to-,thin,black}]
\tikzset{treatment/.style={dashed}}

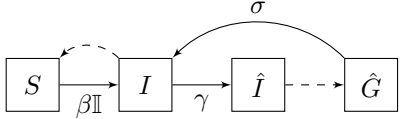
\begin{figure} 
\begin{center}
\begin{tikzpicture}[node distance=1.5cm,auto,>=latex']
\node [int] (S) {$S$};
\node [int] (A) [right of=S] {$I$};

\node [int] (AE) [right of=A] {$\hat{I}$};
\node [int] (E) [right of=AE] {$\hat{G}$};

\path[->] (S) edge node[below] {$\beta \mathbb{I}$} (A);
\draw[->] (A) edge node[below] {$\gamma$} (AE);
\draw[->, treatment] (AE) to (E);

\draw[->, treatment] (A) to[out=135, in=45] (S);

\draw[->] (E) to[out=135,in=45] node[above] {$\sigma$} (A);
\end{tikzpicture}
\end{center}
\caption{The most basic scabies model. The probability of contact and infectiousness is represented by $\beta$, the sexual maturity by $\gamma$ and the time to eggs hatching by
$\sigma$. The total number of infectious individuals is given by $\mathbb{I} = X_I + X_{\hat{I}}$. The solid arrows represent the transitions that occur naturally,
while the dashed arrows represent the effect of a treatment event.}
\label{fig:model1}
\end{figure}

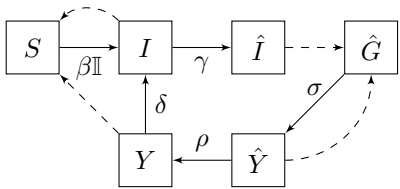
\begin{figure} 
\begin{center}
\begin{tikzpicture}[node distance=1.5cm,auto,>=latex']
\node [int] (S) {$S$};
\node [int] (A) [right of=S] {$I$};

\node [int] (AE) [right of=A] {$\hat{I}$};
\node [int] (E) [right of=AE] {$\hat{G}$};

\node[int] (Y) [below of=A]{$Y$};
\node[int] (YHat) [below of=AE] {$\hat{Y}$};

\path[->] (S) edge node[below] {$\beta \mathbb{I}$} (A);
\draw[->] (A) edge node[below] {$\gamma$} (AE);
\draw[->, treatment] (AE) to (E);

\draw[->, treatment] (A) to[out=135, in=45] (S);

\draw[->] (E) to node[above] {$\sigma$} (YHat);
\draw[->] (YHat) to node[above] {$\rho$} (Y);
\draw[->] (Y) to node[right] {$\delta$} (A);

\draw[->, treatment] (Y) to (S);
\draw[->, treatment] (YHat) to[out=-0, in=-90] (E);
\end{tikzpicture}
\end{center}
\caption{A scabies model incorporating the life cycle of the mite. The
parameter $\delta$ represents rate at which an individual progresses from having only young mites to having at least one mature mite, while $1/\rho$
represents how long until \emph{all} mites hatch. The total number of infectious individuals is given by $\mathbb{I} = X_I + X_{\hat{I}}$}.
\label{fig:model2}
\end{figure}

The maturation of the mite is modelled through the introduction of two new states, $Y$ and $\hat{Y}$, and is represented in Figure \ref{fig:model2}.
Continuing the notation used thus far, the $Y$ state represents individuals who have only young mites, and no eggs, while the $\hat{Y}$ state represents individuals who have young mites with eggs. Now, the likelihood of a successful second treatment has increased, as individuals are `egg-free' while in both the $I$ and $Y$ states. However, an early second treatment will not clear infestation, as individuals may still be carrying unhatched eggs (state $\hat{Y}$). Despite these improvements, the model does not capture any difference in relative susceptibility for subsequent infections, differences in host response between the first and subsequent infections including possible changes in relative infectiousness, nor distinguish endogenous from exogenous reinfection, all of which are well established phenomena of scabies infection as discussed in Section \ref{sec:biology}. Also, this model does not consider the time it takes mites to find an appropriate partner and mate once sexually mature.

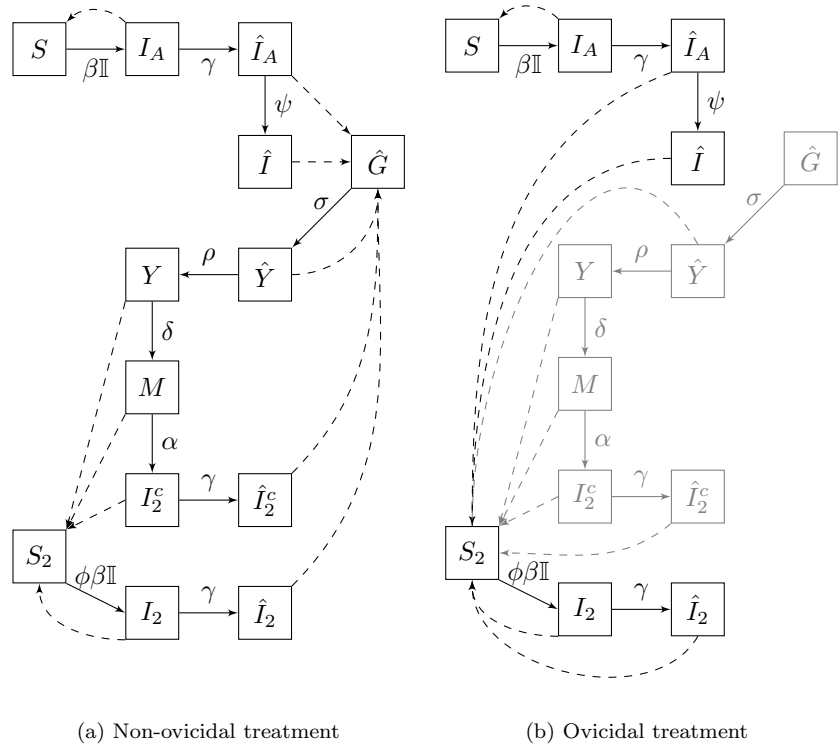
\begin{figure} 
\begin{center}
\begin{tabular}{cc}
\begin{tikzpicture}[node distance=1.5cm,auto,>=latex']
\node [int] (S) {$S$};

\node [int] (IA) [right of=S] {$I_A$};
\node [int] (IAHat) [right of=IA] {$\hat{I}_A$};

\node [int] (IHat) [below of=IAHat] {$\hat{I}$};
\node [int] (E) [right of=IHat] {$\hat{G}$};

\node[int] (YHat) [below of=IHat] {$\hat{Y}$};
\node[int] (Y) [left of=YHat]{$Y$};
\node[int] (M) [below of=Y]{$M$};

\node (D1) [below of=S] {};
\node[int] (I2) [below of=M] {$I_2^c$};
\node[int] (I2Hat) [right of=I2] {$\hat{I}_2^c$};

\node[int] (I2New) [below of=I2] {$I_2$};
\node[int] (I2HatNew) [below of=I2Hat] {$\hat{I}_2$};

\node (D2) at ($(I2)!0.5!(I2New)$) {};
\node[int] (S2) [left of=D2] {$S_2$};

\path[->] (S) edge node[below] {$\beta\mathbb{I}$} (IA);
\draw[->] (IA) edge node[below] {$\gamma$} (IAHat);
\draw[->] (IAHat) edge node {$\psi$} (IHat);
\draw[->, treatment] (IAHat) to (E);
\draw[->, treatment] (IHat) to (E);

\draw[->, treatment] (IA) to[out=135, in=45] (S);

\draw[->, treatment] (Y.south west) to (S2.north east);
\draw[->, treatment] (I2.west) to (S2.north east);
\draw[->, treatment] (I2Hat) to[in=-90] (E);
\draw[->, treatment] (I2HatNew) to[in=-90] (E);
\draw[->, treatment] (I2New.south west) to[out=180, in=-90] (S2.south);
\draw[->, treatment] (M.south west) to (S2.north east);

\draw[->] (S2.south east) to node[above]{$\phi\beta\mathbb{I}$} (I2New.west);
\draw[->] (I2) to node{$\gamma$} (I2Hat);
\draw[->] (I2New) to node{$\gamma$} (I2HatNew);

\draw[->] (E) to node[above] {$\sigma$} (YHat);
\draw[->] (YHat) to node[above] {$\rho$} (Y);
\draw[->] (Y) to node[right] {$\delta$} (M);
\draw[->] (M) to node[right] {$\alpha$} (I2);

\draw[->, treatment] (YHat) to[out=0, in=-90] (E);
\end{tikzpicture}
&
\raisebox{-1cm}{
\begin{tikzpicture}[node distance=1.5cm,auto,>=latex']
\node [int] (S) {$S$};

\node [int] (IA) [right of=S] {$I_A$};
\node [int] (IAHat) [right of=IA] {$\hat{I}_A$};

\node [int] (IHat) [below of=IAHat] {$\hat{I}$};
\node [int, color=gray] (E) [right of=IHat] {$\hat{G}$};

\node[int, color=gray] (YHat) [below of=IHat] {$\hat{Y}$};
\node[int, color=gray] (Y) [left of=YHat]{$Y$};
\node[int, color=gray] (M) [below of=Y] {$M$};

\node (D1) [below of=S] {};
\node[int, color=gray] (I2) [below of=M] {$I_2^c$};
\node[int, color=gray] (I2Hat) [right of=I2] {$\hat{I}_2^c$};

\node[int] (I2New) [below of=I2] {$I_2$};
\node[int] (I2HatNew) [below of=I2Hat] {$\hat{I}_2$};

\node (D2) at ($(I2)!0.5!(I2New)$) {};
\node[int] (S2) [left of=D2] {$S_2$};

\path[->] (S) edge node[below] {$\beta\mathbb{I}$} (IA);
\draw[->] (IA) edge node[below] {$\gamma$} (IAHat);
\draw[->] (IAHat) edge node {$\psi$} (IHat);
\draw[->, treatment] (IAHat.south west) to[out=200, in=90] (S2.north);
\draw[treatment] (IHat.west) to[out=180, in=90, looseness=1] (S2);

\draw[->, treatment] (IA) to[out=135, in=45] (S);

\draw[->, treatment, color=gray] (Y.south west) to (S2.north east);
\draw[->, treatment, color=gray] (I2.west) to (S2.north east);
\draw[->, treatment, color=gray] (I2Hat.south west) to[out=-140, in=0] (S2.east);
\draw[->, treatment] (I2HatNew.south) to[out=-120, in=-90] (S2);
\draw[->, treatment] (I2New.south west) to[out=180, in=-90] (S2.south);
\draw[->, treatment, color=gray] (M.south west) to (S2.north east);

\draw[->] (S2.south east) to node[above]{$\phi\beta\mathbb{I}$} (I2New.west);
\draw[->, color=gray] (I2) to node{$\gamma$} (I2Hat);
\draw[->] (I2New) to node{$\gamma$} (I2HatNew);

\draw[->, color=gray] (E) to node[above] {$\sigma$} (YHat);
\draw[->, color=gray] (YHat) to node[above] {$\rho$} (Y);
\draw[->, color=gray] (Y) to node[right] {$\delta$} (M);
\draw[->, color=gray] (M) to node[right] {$\alpha$} (I2);

\draw[treatment, color=gray] (YHat.north) to[out=120, in=90, looseness=1.4] (S2);
\end{tikzpicture}}
\\
{\footnotesize (a) Non-ovicidal treatment} & {\footnotesize (b) Ovicidal treatment}
\end{tabular}
\end{center}
\caption{A scabies model incorporating the life cycle of the mite, a
subsequent infection phase, and tracking exogenous and endogenous infections
separately with (a) non-ovicidal treatment and (b) ovicidal treatment. The expression for $\mathbb{I}$ is given in Equation \eqref{eqn:I}. The parameter $\psi$ represents the rate at which symptoms to develop,
and $\phi$ represents the relative susceptibility for subsequent infections. The grey states in (b) are inaccessible for realistic initial conditions.}
\label{fig:model3}
\end{figure}

These features are included in our full model (Figure
\ref{fig:model3}). An asymptomatic period has been introduced after the first
infection (denoted by the $I_A$ and $\hat{I}_A$ states). We assume that
repeated asymptomatic infections are unlikely, except in the instance of very
early treatment. Therefore, treatment from the $\hat{I}_A$ state is similar to that
for any other stage of infection, transitioning to state $\hat{G}$ under a non-ovicidal treatment scheme. Further, secondary infections have been
separated into two states. Continuing, or \emph{endogenous} reinfection progresses first through an adult mite state ($M$), and then through $I_2^c$ and $\hat{I}_2^c$, while new or \emph{exogenous} reinfection
occurs through $I_2$ and $\hat{I}_2$. This model includes all the biologically
relevant features of the mite identified in Section \ref{sec:biology}, including maturation and the
production of eggs.

The model can be considered as a continuous-time Markov chain, with state space,
\begin{equation}
\Omega = \{S, I_A, \hat{I}_A, \hat{I}, \hat{G}, \hat{Y}, Y, M, S_2, I_2^c, \hat{I}_2^c, I_2, \hat{I}_2\}.
\end{equation}

Let the number of individuals in state $i \in \Omega$ at time $t$ be $X_i(t)$, and $\vect{X}(t) = \{X_i(t)\}$. All events then correspond to removing an individual from a given state and placing a new individual in the destination state; thus, the total number of individuals in the population remains fixed. For simplicity, we suppress the explicit dependence on time from the state of the Markov chain. For example, an initial infection event from the state $S$ to the state $I_A$ is given by the state transition,
$$\left( X_S, X_{I_A}\right) \rightarrow \left( X_S - 1, X_{I_A} +1 \right).$$

The dynamics of the model can be divided into two sections --- the \emph{natural} transitions and the \emph{treatment} transitions.

\subsection{Natural Transitions}
The natural transitions of the model consist of all the transitions which occur without any treatment or intervention. We assume that the population mixes homogeneously, and that contact rates are frequency dependent. Thus, the initial infection transition,
$$\left( X_S, X_{I_A}\right) \rightarrow \left( X_S - 1, X_{I_A} +1 \right),$$
occurs at rate $$\frac{\beta}{N-1} X_S \mathbb{I}(t),$$ where $N$ is the number of individuals in the population and
\begin{equation} \label{eqn:I}
\mathbb{I}(t) = X_{I_A} + X_{\hat{I}_A} + X_{\hat{I}} + X_{I_2^c} + X_{\hat{I}_2^c} + X_{I_2} + X_{\hat{I}_2},
\end{equation}
is the total number of infectious individuals at time $t$. Similarly, for subsequent exogenous infections, the transition is,
$$\left( X_{S_2}, X_{I_2} \right) \rightarrow \left( X_{S_2} - 1, X_{I_2}+1 \right),$$
which occurs at rate $$\frac{\phi\beta}{N-1} X_{S_2} \mathbb{I}(t),$$
where $\phi \in [0, 1]$ is the relative susceptibility to secondary infections.

The transition from having pregnant mites only to adult mites and eggs ($X_{I} \rightarrow X_{\hat{I}}$) occurs at the rate at which adult mites lay eggs, $\gamma$. The per-individual rate of developing symptoms is given by $\psi$. The per-individual rate for eggs to begin hatching is given by $\sigma$, and the per-individual rate until all eggs have hatched (once one has hatched) is given by $\rho$. The rate at which the mites have completed development and become adults is $\delta$, and the rate at which these mites form breeding pairs is $\alpha$.

Finally, the model allows for human population turnover through births and deaths. All new births enter the $S$ class, while the per-capita death rate out of each class is given by $\mu$, independent of disease status. Throughout, we assume that births and deaths are balanced.

\subsection{Treatment Transitions}
Continuous background treatment is modelled using a constant per-individual rate, $\tau$. Mass drug administration starting at time $\bar{t}$, whereby treatments are distributed at a fixed rate for a specified period of time, $[\bar{t}, \bar{t}+\kappa)$, is modelled as a constant (\emph{not} per-individual) rate out of each infected class. This rate is given by,
$$\omega_{X_i}(t) = \begin{cases}
\frac{\eta X_i(\bar{t})}{\kappa}\mathcal{I}\{X_i \geq 0\}, & \text{ for } t \in [\bar{t}, \bar{t}+\kappa) \\
0, &\text{ otherwise} \end{cases}$$
for each infected state, where $\eta \in [0, 1]$ is the effective coverage of the MDA, incorporating both population coverage and efficacy, and $\kappa$ is the duration of the MDA. This rate is combined with an indicator function, $\mathcal{I}\{X_i\geq 0\}$, in order to ensure each $X_i(t) \geq 0$. An MDA treatment rate of this form ensures that the expected number of individuals treated over the period of the MDA is equal to the effective coverage of the MDA multiplied by the number of infected individuals when the MDA is commenced.

Consider firstly treatment which is non-ovicidal. The dynamics of this treatment are depicted in Figure \ref{fig:model3} (a) with the dotted arrows. For individuals who have both eggs and mites, the first treatment will move individuals to the eggs only state, $\hat{G}$, while an optimally timed second treatment would move an individual to the $S_2$ state (from state $Y, M$ or $I_2^c$). Treatment at a non-optimal time may simply move an individual straight back to the $\hat{G}$ state (from state $\hat{Y}$ if too early; or from $\hat{I}_2^c$ if too late), and so it is important to consider the time between interventions, or the \emph{intervention interval}, for non-ovicidal scabies treatments.

Comparatively, if treatment were ovicidal (Figure \ref{fig:model3} (b)), then the destination state for each treatment event will be state $S_2$, except from the state $I_A$, for which $S$ remains the destination due to an assumption of no acquisition of immunity while experiencing asymptomatic infection. It is worth noting that under an ovicidal treatment scheme, the states $\hat{G}, \hat{Y}, Y, M, I_2^c$ and $\hat{I}_2^c$, are ephemeral, and so play no role in the equilibrium solutions of the system.

The size of the state space, $|\Omega|$, is prohibitively large to allow numerical solutions to the forward equations to be found for anything but small values of $N$. However, the process may be simulated using the so-called Gillespie algorithm \cite{Gillespie1977}. The mean behaviour of the process can be explored using a (deterministic) mean-field approximation. The full mean-field approximation, $\vect{x}(t)$, is detailed in \ref{app:meanfield}.

We use the \texttt{ode45} routine in MATLAB to calculate numerical solutions for this mean-field approximation. The parameter values used for this work are detailed in Table \ref{tab:params}, with further details provided in \ref{app:foi}. Many of the values relating to the natural history of the mite have been determined from laboratory experiments \cite{Arlian1988}, with the exception of the mean time for a mite to lay eggs, which is controlled by the parameter $\gamma$. We choose $\gamma$ based on estimates for the mean time to the second generation of mites appearing post-infection \cite{Mellanby1944}.

\begin{sidewaystable}
\centering
\begin{tabular}{c|c|c|c|c}
Parameter & Meaning & Value$^{-1}$ (days) & Reference & Notes \\ \hline
$\beta$ & Probability of contact and infectiousness & 63 & \cite{Kearns2013}, \cite{Romani2015a} &\\
$\phi$ & Relative susceptibility to subsequent infections & 0.5 & &  Entire range of $\phi \in [0, 1]$ explored in \ref{app:reduced-susceptibility}\\
$\alpha$ & Rate of mating & 15 & \cite{Mellanby1944} & \\
$\gamma$ & Rate of egg laying & 0.5 & \cite{UniversityofSydney2016} \\
$\psi$ & Rate of symptoms developing & 30 & \cite{Mellanby1944} & \\
$\sigma$ & Rate of egg hatching & 2 & \cite{Arlian1988} & \\
$\rho$ & Rate for mites to become nymphs & 5  & \cite{Arlian1988} & \\
$\delta$ & Rate of maturation & 5 & \cite{Arlian1988} &\\
$\tau$ & Rate of background treatment (non-ovicidal) & 57.13 & \cite{Kearns2013} & Detailed in \ref{app:foi}\\
& \phantom{Rate of background treatment} (ovicidal) & 173.7 & \cite{Kearns2013} & Detailed in \ref{app:foi}\\
$\mu$ & Death rate & 18250 & & 50 year life expectancy \\
\end{tabular}
\caption{Parameter values (presented as the inverse) used in the model. All parameters are given in days.} \label{tab:params}
\end{sidewaystable}

\subsection{Objective Function}
In order to determine optimal intervention intervals, we formulate a mathematical optimisation problem. The objective function is chosen to be the minimum proportion of the population who are carrying eggs over the time period of interest. That is,
\begin{equation} \label{eqn:optimisationproblem}
\begin{aligned}
& \underset{t_1 \dots t_n} {\text{minimize}}  & & \min(e(t)) \\
& \text{subject to} & & 0 < t_1 < \dots < t_n,
\end{aligned}
\end{equation}
where,
$$e(t) = x_{\hat{I}_A} + x_{\hat{I}} + x_{\hat{G}} + x_{\hat{Y}} + x_{\hat{I}_2^c} + x_{\hat{I}_2}$$
is the proportion of the population who have eggs at time $t$, and  $t_1, \dots, t_n$ are the times at which a treatment round in the MDA occurs.

The first intervention time, $t_1$, is entirely arbitrary, so in practice we fix $t_1$ and the remaining intervention times, $t_2, \dots, t_n$ are determined. Solutions are calculated numerically using the \texttt{fmincon} routine in MATLAB.

Note that the problem stated in Equation \eqref{eqn:optimisationproblem} utilises the mean-field approximation, $\vect{x}(t)$. However, the same solution could also be obtained by formulating the problem using the mean of the Markov chain, $\mathbb{E}[\vect{X}(t)]$, as the only difference between the problem formulations is the constant $1/N$.

\section{Results}
\begin{figure}
\centering
\includegraphics[scale=0.7]{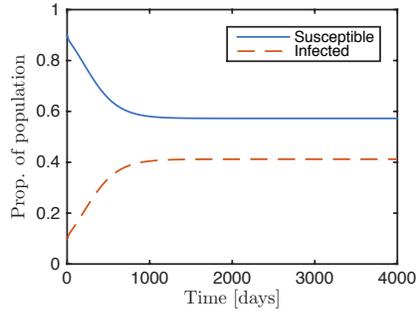}
\caption{Overall dynamics of the system including background treatment before reaching equilibrium using the parameters in Table \ref{tab:params}.} \label{fig:dynamics}
\end{figure}

First, we establish the equilibrium dynamics in the absence of an MDA. Using the mean-field approximation, the model can be divided into three classes: susceptible, infected and latent. The susceptible class consists of the states $S$ and $S_2$, while the infected class consists of $\Omega \backslash \{S, S_2, \hat{G}\}$. Note that we make an important distinction between an infected and an \emph{infectious} state. Only individuals who are harboring pregnant mites, and thus in states $I, \hat{I}, I_2^c, \hat{I}_2^c, I_2$ and $\hat{I}_2$ are classified as infectious. Individuals who are harboring \emph{any} mites are considered infected. That is, all infectious states as well as states $Y, \hat{Y}$ and $M$ are considered infected. The latent class consists of only the $\hat{G}$ state, as left untreated, individuals will inevitably progress to be infected. However, the $\hat{G}$ state is an \emph{untreatable} latent state under a non-ovicidal treatment scheme, which is different to the latent phases incorporated into other disease models \cite{Anderson1992,Keeling2008}. These three classes form SIS-like dynamics, as shown in Figure \ref{fig:dynamics}.

\begin{figure}
\centering
\includegraphics[scale=0.6]{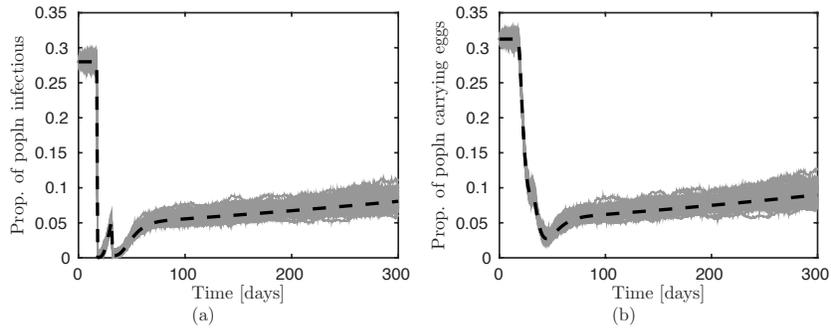}
\caption{50 realisations of the Markov model (grey lines), using the Gillespie algorithm, compared with the mean-field approximation (black dotted line) with two optimally timed non-ovicidal interventions on (a) the proportion of infected individuals and (b) the proportion of the population with eggs, using the parameters in Table \ref{tab:params}.} \label{fig:comparison}
\end{figure}

Before studying the actions of mass drug administration, we provide numerical justification for the mean-field approximation, $\vect{x}(t)$, to the Markov chain, $\vect{X}(t)$. Figure \ref{fig:comparison} shows that the system of ordinary differential equations given in \ref{app:meanfield} approximates the mean behaviour of the system well, even in the event of an MDA. As such, we will focus primarily on the results from the mean-field approximation.

Having established non-MDA equilibria and the utility of the mean-field approximation, we now investigate the impact of interventions on the system. One crucial parameter here is the proportion of the infected population who are successfully treated during a mass drug administration, $\eta$. For the purpose of exploring model behaviour under an MDA with non-ovicidal treatment, we assume that $100\%$ of the population is treated $100\%$ effectively. While this is not achievable in practice, this provides a theoretical \emph{best-case} scenario for an MDA.

\subsection{Intervention Intervals}
\begin{figure}
\centering
\includegraphics[scale=0.8]{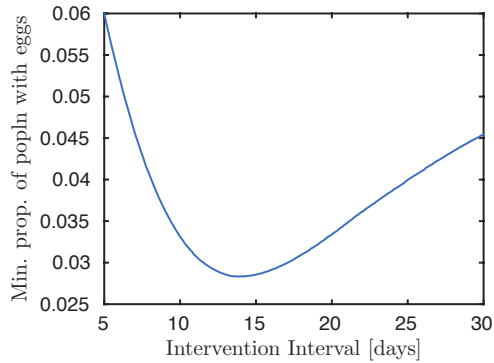}
\caption{Impact of varying inter-intervention intervals for a $100\%$ effective MDA using a non-ovicidal treatment, using the parameter values in Table \ref{tab:params}, for an MDA with two rounds of treatment.} \label{fig:interinterventionintervals}
\end{figure}

Recall that when considering non-ovicidal interventions, it is clear that at least two treatments are needed. We investigate the impact of the time between these treatments in the event of an MDA. A local minimum for the optimisation problem in Equation \eqref{eqn:optimisationproblem} is identified in Figure \ref{fig:interinterventionintervals}. If the second treatment occurs too early, then not all the eggs will have hatched (state $X_{\hat{Y}}$), leading to endogenous reinfection, while if the treatment occurs late, newly hatched mites will have matured, mated and produced new eggs (state $X_{\hat{I}_2^c}$). Solving the optimisation problem numerically using the MATLAB routine \texttt{fmincon} gives an optimal intervention interval of 13.94 days.

This intervention interval is closely linked to the modelled life-cycle of the mite. With the parameter values in Table \ref{tab:params}, the mean time to adult mites reappearing (assuming no second treatment occurs) is 12 days. Intuitively, the inter-intervention interval should be long enough so that all of the previous eggs have hatched, but short enough so that the number of individuals infested with eggs is small. This aligns closely with the calculated optimal intervention interval of approximately 14 days.

Returning to Figure \ref{fig:comparison}, it shows the dynamics of the system using two optimally timed MDA interventions using a non-ovicidal treatment. It is clear that even with an MDA which is maximally effective, two doses is not sufficient for eradication of scabies from the population. In less than one year, prevalence increases above $5\%$ and after approximately 10 years (not pictured) the prevalence has returned to baseline levels. Even though the prevalence of infectious individuals was reduced to very low levels (Figure \ref{fig:comparison} (a)), the proportion of the population with eggs remains high (Figure \ref{fig:comparison} (b)), and so endogenous reinfection and rebound is inevitable.

\begin{figure}
\centering
\includegraphics[scale=0.8]{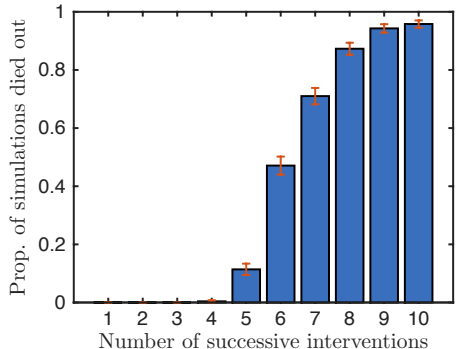}
\caption{Mean, $\bar{x}$, of 1000 simulations of the Markov chain, $\vect{X}(t)$, and $95\%$ confidence intervals (calculated as $\bar{x} \pm 1.96\times s/\sqrt{N}$, where $s$ is the standard deviation of the sample) which experienced die-out for varying numbers of successive optimally timed non-ovicidal MDA treatments, using the parameter values in Table \ref{tab:params} and a population size of $N=2000$.} \label{fig:noofinterventions}
\end{figure}

Having demonstrated that two non-ovicidal interventions is not sufficient for eradication of scabies in this model, we now consider how many interventions would be required for eradication with non-ovicidal treatment. Figure \ref{fig:noofinterventions} shows the proportion of 1000 Gillespie simulations which resulted in eradication of scabies for a population size of $N=2000$ (a typical size for a high-prevalence community in Northern Australia). In the case of having ten optimally timed interventions inside the MDA, $93\% (\pm 0.8\%)$ of the simulations showed eradication of scabies. This further demonstrates the difficulties with achieving eradication of scabies using a non-ovicidal treatment. Even with ten optimally timed interventions, which would require drastically more resources than are currently available, eradication is by no means guaranteed. These results have been generated under the assumption that the relative susceptibility to secondary infections, $\phi=0.5$ (\ref{tab:params}). \ref{app:reduced-susceptibility} shows that these findings are robust to alternative assumptions on the level of relative susceptibility to secondary infections, $\phi$.

\subsection{Mass Drug Administration with an ovicidal treatment}
Thus far, we have considered only non-ovicidal treatment for scabies. Figure \ref{fig:model3} (b) represents the dynamics using an ovicidal treatment. We now investigate the effects of an MDA using an ovicidal treatment.

When using an ovicidal treatment with the parameters in Table \ref{tab:params}, the prevalence of scabies is zero under only background treatment (using $\tau = 1/57.13$) due to the increased effectiveness of this treatment and assumed coverage level. Accordingly, we recalibrate the model to give the same endemic prevalence as used in our non-ovicidal example of $28\%$, achieved by setting $\tau = 1/173.7$.

\begin{figure}
\centering
\begin{tabular}{cc}
\includegraphics[scale=0.65]{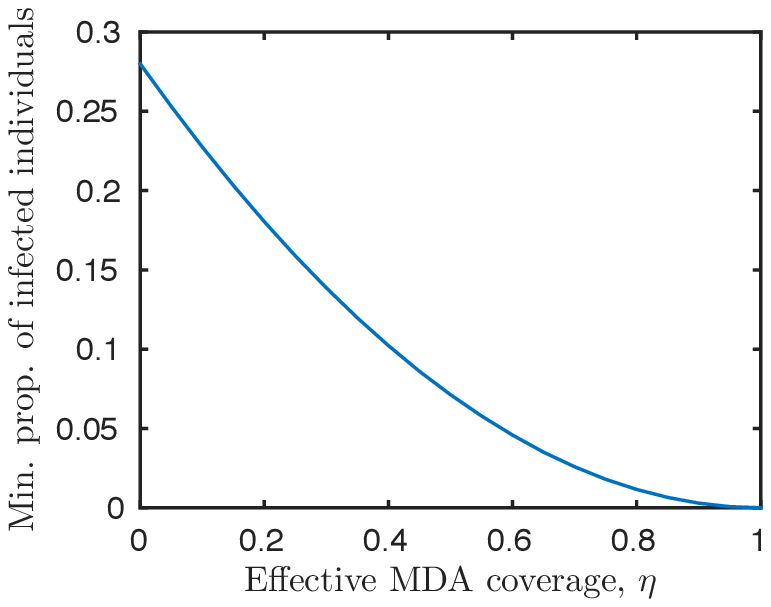} \hspace{20mm}& \includegraphics[scale=0.65]{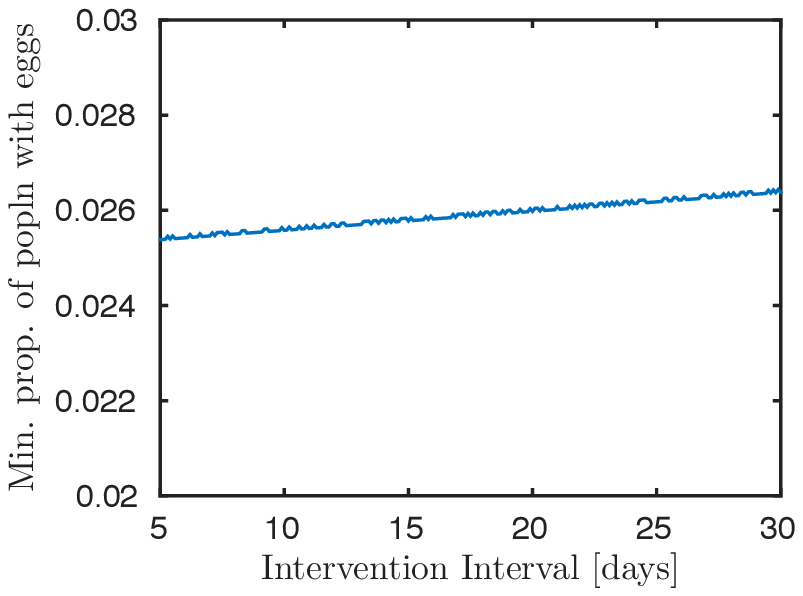} \\
{\footnotesize (a)} & {\footnotesize (b)}
\end{tabular}
\caption{(a) Comparison of the coverage of an MDA using an ovicidal treatment on the minimum proportion of infected individuals achieved and (b) impact of varying the inter-intervention interval using an ovicidal treatment on the minimum proportion of the population that has eggs using the parameter values in Table \ref{tab:params}, and $\tau = 1/173.7$ and an MDA effective coverage of $\eta = 0.7$.} \label{fig:ovicidalmdacomparison}
\end{figure}

Clearly, a maximally effective (i.e. $100\%$ coverage, $100\%$ effective) MDA for an ovicidal treatment will eradicate infection in the population. Figure \ref{fig:ovicidalmdacomparison} (a) shows how variation in the coverage of an MDA influences the outcome. Note that coverage and compliance are confounded in our model. Unsurprisingly, between the extremes, the minimum proportion of infected individuals is monotonically decreasing.

Choosing an MDA coverage of $70\%$ and allowing for a second round of mass drug administration, Figure \ref{fig:ovicidalmdacomparison} (b) shows the minimum proportion of the population who are carrying eggs as a function of the interval between treatments. This result suggests a `sooner is better' approach to a follow up round of MDA when using an ovicidal treatment scheme. However, we emphasise this result arises from coverage and compliance being subsumed into the same measure and so all individuals are assumed to be equally likely to be administered treatment in the second course.

\section{Discussion}
We have developed a biologically informed mathematical model of scabies infestations in a population which explicitly accounts for the multiple life stages of the mite and the presence of eggs. While there have been other models of scabies proposed \cite{Bhunu2013,Gilmore2011}, they have not adequately captured the critical features of the life-cycle of the mite. This model has provided a framework with which to explore the different consequences of ovicidal and non-ovicidal treatment strategies. Crucially, the model is able to qualitatively reproduce the recrudescence of infection post mass drug administration that has been observed in practice \cite{Carapetis1997,Currie2000}, despite the prevalence of scabies becoming very small.

Our analysis has demonstrated that even under the assumption of $100\%$ coverage and efficacy for a non-ovicidal treatment, rebound of infection levels is inevitable. Comparatively, a single dose of an ovicidal treatment with full compliance and effectiveness is sufficient to eradicate infection. In the event of an MDA with a non-ovicidal treatment, we have shown that the intervention interval should be approximately two weeks, a number a closely linked to the life-cycle of the mite.

Furthermore, when considering the effective coverage with ovicidal treatment, we demonstrated that the relationship between effective coverage and short-term success is monotonic. If the issue with ovicidal drugs is one of compliance, then it is a clear limitation of this model that compliant and non-compliant individuals are not separated. This results in ovicidal treatment impact being generally overestimated.

The model formulated here forms a basis for modelling scabies infection in a population, but is by no means exhaustive. For example, we have not considered reduced levels of infectiousness for subsequent infections or any form of immunity. These features could potentially be included alongside an age structured model once data becomes available \cite{Romani2015,Romani2015b}.

This model has provided the first mathematical insight into the limited long-term success of MDA treatment of scabies in areas of endemic prevalence. By using biologically relevant parameters, best-case MDA scenarios and issues surrounding compliance have been explored, demonstrating the difficulty that is associated with control of this infection.

\section{Acknowledgements}
M.~Lydeamore is supported by an Australian Postgraduate Award; J.~McVernon is supported by an NHMRC Career Development Fellowship (CDF1061321); J.~M.~McCaw is supported by an ARC Future Fellowship (FT110100250). D.~Regan is supported by an NHMRC Program Grant (APP1071269); S.~Y.~C.~Tong is supported by an NHMRC Career Development Fellowship (CDF1065736); We thank the NHMRC Centre for Research Excellence in Infectious Diseases Modelling to Inform Public Health Policy (1078068). This work is supported by an NHMRC Project Grant titled `Optimising intervention strategies to reduce the burden of Group A Streptococcus in Aboriginal Communities' (APP1098319).

\section{Bibliography}
\bibliographystyle{abbrv}
\bibliography{library}

\appendix
\section{Transitions for model dynamics}
\subsection{Model A}
The set of natural transitions is as follows: the infection transition is,
\begin{equation} \label{eqn:model1eqn1}
(X_S, X_I) \rightarrow (X_S-1, X_I+1) \text{ at rate } \beta X_S (X_I + X_{\hat{I}}),
\end{equation}
while the egg-laying transition is,
\begin{equation}
(X_I, X_{\hat{I}}) \rightarrow (X_I-1, X_{\hat{I}}+1) \text{ at rate } \gamma X_I,
\end{equation}
and the egg-hatching transition (post-treatment) is,
\begin{equation}
(X_{\hat{G}}, X_I) \rightarrow (X_{\hat{G}}-1, X_I+1) \text{ at rate } \sigma X_{\hat{G}}.
\end{equation}
The set of treatment transitions is,
\begin{equation}
(X_{\hat{I}}, X_{\hat{G}}) \rightarrow (X_{\hat{I}}-1, X_{\hat{G}}+1) \text{ at rate } \tau X_{\hat{I}},
\end{equation}
when an individual has adult mites and eggs on them, and
\begin{equation} \label{eqn:model1eqnlast}
(X_I, X_S) \rightarrow (X_I - 1, X_S + 1) \text{ at rate } \tau X_I,
\end{equation}
for an individual carrying no eggs.

\subsection{Model B}
The set of natural transitions is as follows: the infection transition is,
\begin{equation} \label{eqn:model2eqn1}
(X_S, X_I) \rightarrow (X_S-1, X_I+1) \text{ at rate } \beta X_S (X_I + X_{\hat{I}}),
\end{equation}
while the egg-laying transition is,
\begin{equation}
(X_I, X_{\hat{I}}) \rightarrow (X_I-1, X_{\hat{I}}+1) \text{ at rate } \gamma X_I,
\end{equation}
and the egg-hatching transition (post-treatment) is,
\begin{equation}
(X_{\hat{G}}, X_{\hat{Y}}) \rightarrow (X_{\hat{G}}-1, X_{\hat{Y}}+1) \text{ at rate } \sigma X_{\hat{G}}.
\end{equation}
The transition where the last egg hatches, and only young mites remain is,
\begin{equation}
(X_{\hat{Y}}, X_Y) \rightarrow (X_{\hat{Y}} -1, X_Y + 1) \text{ at rate } \rho X_{\hat{Y}},
\end{equation}
and the maturation of the first mite is,
\begin{equation}
(X_Y, X_I) \rightarrow (X_Y - 1, X_I + 1) \text{ at rate } \delta X_Y.
\end{equation}
The set of treatment transitions is,
\begin{equation}
(X_{\hat{I}}, X_{\hat{G}}) \rightarrow (X_{\hat{I}}-1, X_{\hat{G}}+1) \text{ at rate } \tau X_{\hat{I}},
\end{equation}
when an individual has adult mites and eggs on them,
\begin{equation}
(X_{\hat{Y}}, X_{\hat{G}}) \rightarrow (X_{\hat{Y}} - 1, X_{\hat{G}} + 1) \text{ at rate } \tau X_{\hat{Y}},
\end{equation}
for when an individual has young mites and eggs and
\begin{equation} \label{eqn:model2eqnlast}
(X_I, X_S) \rightarrow (X_I - 1, X_S + 1) \text{ at rate } \tau X_I,
\end{equation}
for an individual carrying no eggs.

\section{Mean Field Approximation} \label{app:meanfield}
The full set of ordinary differential equations which make up the mean field approximation with non-ovicidal treatment is,
\begin{align*}
\dot{x}_S &= -\beta x_S \mathbb{I}(t) + \tau x_{I_A} + \mu (1-x_S) + \omega_{x_{I_A}}(t), \\
\dot{x}_{I_A} &= \beta x_S \mathbb{I}(t) - (\gamma + \tau + \mu) x_{I_A} - \omega_{x_{I_A}}(t), \\
\dot{x}_{\hat{I_A}} &= \gamma x_{I_A} - (\psi + \tau + \mu) x_{\hat{I_A}} - \omega_{x_{\hat{I}_A}}(t), \\
\dot{x}_{\hat{I}} &= \psi x_{\hat{I_A}} - (\tau + \mu) x_{\hat{I}} - \omega_{x_{\hat{I}}}(t), \\
\dot{x}_{\hat{G}} &= \tau (x_{\hat{I}_A} + x_{\hat{I}} + x_{\hat{Y}} + x_{\hat{I}_2^c} + x_{\hat{I}_2}) - (\sigma + \mu) x_{\hat{G}} \\
&\quad+ \omega_{x_{I_A}}(t) + \omega_{x_{\hat{I}}}(t) + \omega_{x_{\hat{Y}}}(t) + \omega_{x_{\hat{I}_2^c}}(t) + \omega_{x_{\hat{I}_2}}(t) , \\
\dot{x}_{\hat{Y}} &= \sigma x_{\hat{G}} - (\rho + \tau + \mu) x_{\hat{Y}} - \omega_{x_{\hat{Y}}}(t) , \\
\dot{x}_Y &= \rho x_{\hat{Y}} - (\delta + \tau + \mu) x_Y - \omega_{x_{Y}}(t) , \\
\dot{x}_M &= \delta x_Y - (\alpha + \tau + \mu) x_M - \omega_{x_{M}}(t), \\
\dot{x}_{S_2} &= \tau (x_Y + x_M + x_{I_2^c} + x_{I_2}) - \phi\beta x_{S_2} \mathbb{I}(t) - \mu x_{S_2} + \omega_{x_{Y}} + \omega_{x_{I_2^c}}(t) + \omega_{x_{I_2}}(t) , \\
\dot{x}_{I_2^c} &= \alpha x_M - (\gamma +\tau + \mu) x_{I_2^c} - \omega_{x_{I_2^c}}(t) , \\
\dot{x}_{\hat{I}_2^c} &= \gamma x_{I_2^c} - (\tau + \mu) x_{\hat{I}_2^c} - \omega_{x_{\hat{I}_2^c}}(t) , \\
\dot{x}_{I_2} &= \phi\beta x_{S_2} \mathbb{I}(t) - (\gamma + \tau + \mu) x_{I_2} - \omega_{x_{I_2}}(t) , \\
\dot{x}_{\hat{I}_2} &= \gamma x_{I_2} - (\tau + \mu) x_{\hat{I}_2} - \omega_{x_{\hat{I}_2}}(t) ,
\end{align*}
where,
$$\mathbb{I}(t) = x_{I_A} + x_{\hat{I}_A} + x_{\hat{I}} + x_{I_2^c} + x_{\hat{I}_2^c} + x_{I_2} + x_{\hat{I}_2},$$
is the proportion of infected individuals at time $t$.

For ovicidal treatment, the mean field approximation is,
\begin{align*}
\dot{x}_S &= -\beta x_S \mathbb{I}(t) + \tau x_{I_A} + \mu (1-x_S) + \omega_{x_{I_A}}(t), \\
\dot{x}_{I_A} &= \beta x_S \mathbb{I}(t) - (\gamma + \tau + \mu) x_{I_A} - \omega_{x_{I_A}}(t), \\
\dot{x}_{\hat{I_A}} &= \gamma x_{I_A} - (\psi + \tau + \mu) x_{\hat{I_A}}  - \omega_{x_{\hat{I}_A}}(t), \\
\dot{x}_{\hat{I}} &= \psi x_{\hat{I_A}} - (\tau + \mu) x_{\hat{I}} - \omega_{x_{\hat{I}}}(t), \\
\dot{x}_{\hat{G}} &= (\sigma + \mu) x_{\hat{G}}, \\
\dot{x}_{\hat{Y}} &= \sigma x_{\hat{G}} - (\rho + \tau + \mu) x_{\hat{Y}} - \omega_{x_{\hat{Y}}}(t) , \\
\dot{x}_Y &= \rho x_{\hat{Y}} - (\delta + \tau + \mu) x_Y  - \omega_{x_{Y}}(t), \\
\dot{x}_M &= \delta x_Y - (\alpha + \tau + \mu) x_M - \omega_{x_{M}}(t), \\
\dot{x}_{S_2} &= \tau (x_{\hat{I}_A} + x_{\hat{I}} + x_Y + x_M + x_{\hat{Y}} + x_{I_2^c} + x_{\hat{I}_2^c} + x_{I_2} + x_{\hat{I}_2}) - \phi\beta x_{S_2} \mathbb{I}(t) - \mu x_{S_2},  \\
&\quad+ \omega_{x_{I_A}}(t) + \omega_{x_{\hat{I}}}(t) + \omega_{x_{Y}} + \omega_{x_{\hat{Y}}}(t) + + \omega_{x_{I_2^c}}(t) + \omega_{x_{\hat{I}_2^c}}(t) + \omega_{x_{I_2}}(t) + \omega_{x_{\hat{I}_2}}(t)\\
\dot{x}_{I_2^c} &= \delta x_Y - (\gamma +\tau + \mu) x_{I_2^c} - \omega_{x_{I_2^c}}(t), \\
\dot{x}_{\hat{I}_2^c} &= \gamma x_{I_2^c} - (\tau + \mu) x_{\hat{I}_2^c} - \omega_{x_{\hat{I}_2^c}}(t) , \\
\dot{x}_{I_2} &= \phi\beta x_{S_2} \mathbb{I}(t) - (\gamma + \tau + \mu) x_{I_2} - \omega_{x_{I_2}}(t), \\
\dot{x}_{\hat{I}_2} &= \gamma x_{I_2} - (\tau + \mu) x_{\hat{I}_2} - \omega_{x_{\hat{I}_2}}(t)
\end{align*}

\section{Determining probability of contact and infectiousness, and rate of treatment} \label{app:foi}

The mean time to first infection for scabies is 225 days \cite{Kearns2013}. We utilise the common approximation that the \emph{force of infection}, $\lambda$ is equal to the inverse of the mean time to first infection \cite{Keeling2008}. That is,

$$\lambda = \frac{1}{225}.$$

We utilise the relationship,
$$\lambda = \beta \mathbb{I},$$
and take the prevalence of scabies to be $\mathbb{I} = 0.28$ \cite{Romani2015a} for the purposes of this study. This gives,
$$\beta = \frac{\lambda}{\mathbb{I}} = \frac{1}{63}.$$

Considering the system, $\vect{x}(t)$, at equilibrium, and all parameters except $\tau$ and $\phi$ fixed and that there is no natural recovery in this model, the rate of new infections is equal to the combined rate of treatment and death out of each infected state. As the death rate, $\mu$, is much smaller than the per-individual treatment rate, $\tau$. it follows that,

$$\text{total rate of new infections} = \lambda(x_S+\phi x_{S_2}) + \sigma x_{\hat{G}} = \text{total rate of treatment}.$$

The total rate of treatment is,

$$\text{total rate of treatment} = \tau(1-(x_S+x_{S_2}+x_{\hat{G}})),$$
and so,

\begin{equation} \label{eqn:lambdatau}
\lambda(x_S+\phi x_{S_2}) + \sigma x_{\hat{G}} = \tau(1-(x_S+x_{S_2}+x_{\hat{G}})).
\end{equation}

\begin{figure}
\centering
\includegraphics[width=0.5\textwidth]{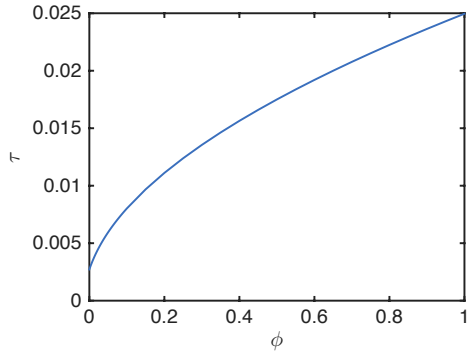}
\caption{Background treatment rate, $\tau$, as a function of the reduction in susceptibility, $\phi$.} \label{fig:tauPhiComparison}
\end{figure}

We assume the prevalence of scabies to be 0.28 \cite{Romani2015a}. That is, $$1-(x_S+x_{S_2}+x_{\hat{G}}) = 0.28.$$ Thus,

\begin{equation} \label{eqn:determinetau}
0.28 \tau = \lambda(x_S+\phi x_{S_2}) + \sigma x_{\hat{G}}.
\end{equation}
with which we cannot uniquely determine $\tau$ or $\phi$. However, as $\phi$ is bounded between 0 and 1, we can explore the possible range for $\tau$ by solving Equation \eqref{eqn:determinetau} numerically. This is shown in Figure \ref{fig:tauPhiComparison}.

\section{Investigation of relative susceptibility to secondary infections} \label{app:reduced-susceptibility}
\begin{figure}
\centering
\includegraphics[width=0.5\textwidth]{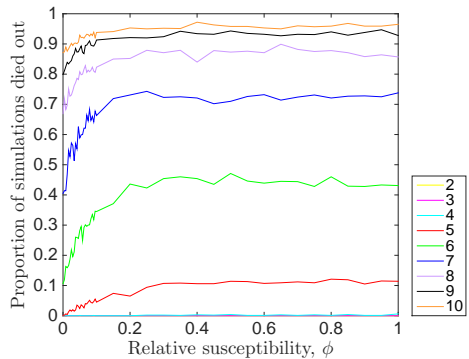}
\caption{Mean, $\bar{x}$, of 1000 simulations of the Markov chain, $\vect{X}(t)$, which experienced die-out for varying numbers of successive optimally timed non-ovicidal MDA treatments and a varying relative susceptibility to secondary infections, $\phi$, with all other parameter values as in Table \ref{tab:params} and a population size of $N=2000$.} \label{fig:reduced-susceptibility}
\end{figure}

Little is known about the level of relative susceptibility to secondary infections for scabies infections. The human infection study for scabies \cite{Mellanby1944} does note an approximately $50\%$ decreased parasite load for individuals experiencing secondary infections, but does not investigate the impact this has on transmission. Here, we consider the entire range of possible values for the relative susceptibility, $\phi$, and measure the influence of this on the probability of eradication. Figure \ref{fig:reduced-susceptibility} shows the estimated probability of eradication following some number of MDA's under a non-ovicidal treatment scheme in a population size of $N=2000$. We observe that when individuals are almost immune to secondary infections, that is when $\phi$ is small, the eradication probability decreases. This decrease is caused by the small rate of background treatment, $\tau$, that is required in the model to maintain the correct prevalence and age of first infection as discussed in \ref{app:foi}. However, it can be seen that the probability of eradication is relatively unchanged for a fixed number of MDA's beyond $\phi = 0.2$. This suggests that the conclusions from this model are robust with respect to the relative susceptibility, $\phi$ (and thus the background treatment rate, $\tau$), unless the relative susceptibility is particularly low.

\end{document}